\documentclass[10pt]{article}
\usepackage[utf8]{inputenc}
\usepackage{amsmath}
\usepackage{graphicx,tabularx}
\usepackage{amsfonts,amsthm,eucal,amssymb,amscd}
\usepackage{color}
\usepackage{epstopdf}
\usepackage{upgreek}
\usepackage{bbold} 
\usepackage{glossaries}

\begin{document}

\newtheorem{theo}{Theorem}[section]
\newtheorem{definition}[theo]{Definition}
\newtheorem{lem}[theo]{Lemma}
\newtheorem{prop}[theo]{Proposition}
\newtheorem{coro}[theo]{Corollary}
\newtheorem{exam}[theo]{Example}
\newtheorem{rema}[theo]{Remark}
\newtheorem{example}[theo]{Example}
\newcommand{\ninv}{\mathord{\sim}} 
\newtheorem{axiom}[theo]{Axiom}

\title{Non-deterministic semantics for quantum states}

\author{{\sc Juan Pablo Jorge}$^{1}$ and {\sc Federico
Holik}$^{2}$}

\date{February 2019}

\maketitle

\begin{center}

\begin{small}
1- Physics Department, University of Buenos Aires, CABA (1428), Argentina; jorgejpablo@gmail.com\\
2- Instituto de F\'{i}sica La Plata, La Plata (1900),
Buenos Aires, Argentina; olentiev2@gamail.com\\
\end{small}
\end{center}

\vspace{1cm}

\begin{abstract}
\noindent In this work, we discuss the failure of the principle of
truth functionality in the quantum formalism. By exploiting this
failure, we import the formalism of N-matrix theory and
non-deterministic semantics to the foundations of quantum mechanics.
This is done by describing quantum states as particular valuations
associated with infinite non-deterministic truth tables. This allows
us to introduce a natural interpretation of quantum states in terms
of a non-deterministic semantics. We also provide a similar
construction for arbitrary probabilistic theories based in
orthomodular lattices, allowing to study post-quantum models using
logical techniques.
\end{abstract}
\bigskip
\noindent

\begin{small}
\centerline{\em Key words: Kochen-Specker thoerem; quantum states;
non-deterministic semantics; truth functionality}
\end{small}

\section{Introduction}\label{s:Introdcution}

According to the principle of truth-functionality or composability
(TFP), the truth-value of a complex formula is uniquely determined
by the truth-values of its subformulas. It is a basic principle of
logic. However, as explained in \cite{Avron-Zamansky}, many
real-world situations involve dealing with information that is
incomplete, vague or uncertain. This is especially true for quantum
phenomena, where only probabilistic assertions about possible events
can be tested in the lab. These situations pose a threat to the
application of logical systems obeying the principle of
truth-functionality in those problems. As is well known, the TFP
fails in many logical systems of interest (see examples
in~\cite{PTF-Examples}).

One possible way to deal with this situation is to relax the TFP.
This is the path followed in~\cite{AvronLev}, where the idea of
non-deterministic computations---borrowed from automata and
computability theory---is applied to evaluations of truth-values of
formulas. This leads to the introduction of non-deterministic
matrices (N-matrices). These are a natural generalization of
ordinary multi-valued matrices, in which the truth-value of a
complex formula can be chosen non-deterministically, out of some
non-empty set of options
\cite{Avron-Zamansky,AvronLev,Avron-Lev-2001b,Avron-Konikowska-2005}.

The Kochen-Specker theorem \cite{KST} is one of the most fundamental
no-go theorems of quantum mechanics, and it holds in many
probabilistic models of interest
\cite{Doring-KSVNA,Svozil-KS,Smith-KS}. It has far-reaching
consequences for the interpretation of the quantum formalism (see
for example \cite{Newton-Olimpia}). In particular, it imposes very
strong restrictions on the possible {physically motivated}
valuations that can be defined over the propositions associated with
quantum models. In this work, we discuss in which sense the
principle of truth functionality {is false} in the quantum formalism
and describe how N-matrices can be used to describe quantum states
and quantum state spaces. This novel perspective opens the door to
new fundamental questions by introducing the possibility of
interpreting quantum probabilities as a particular class of
non-classical valuations.

We also extend our approach to a family of generalized probabilistic
models---including quantum and classical probabilistic ones as
particular cases. The N-matrices associated with classical
probabilistic models are non-deterministic, but they always admit
global classical valuations to the set $\{0,1\}$. These global
valuations will not exist, in general, for non-classical
probabilistic models. The study of post-quantum theories is a very
active field of research nowadays, since it provides an
extraordinary ground for studying the fundamental principles that
underly the quantum formalism (see, for
example,~\cite{Popescu,Cabello2017a,Cabello2016a}). Furthermore, the
study of contextual systems outside the quantum domain
\cite{AcacioContextuality-2015,Aerts,Khrennikiv-Ubiquitous} poses
the question of looking for contextual models which are non-quantum,
but non-classical either. Furthermore, non-standard probabilities
have been used to describe the deviations of classical logic in
decision-making problems. As an example, negative probabilities have
been applied to the study of inconsistent judgments
\cite{AcacioNegative}. Our extension could be useful for studying
contextuality
\cite{Cabello2015,CabelloProposalFor-2010,AcacioContextuality-2015b,Abramsky,deBarros-Holik-Krause-2017}
in quantum mechanics and generalized probabilistic models from a
{novel} logical perspective.

The logic-operational approach to quantum theory dates back to the
1930s, after the contribution of Birkhoff and von Neumann \cite{BvN}
(for further developments of the quantum logical approach,
see~\cite{svozillibro,belcas81,dallachiaragiuntinilibro,piron,aertsdaub1,aertsdaub2,HandbookofQL,mikloredeilibro,Holik-Domenech,Holik
2012,Holik-Zuberman-2013,Fortin 2014,Fortin 2016,Losada 2017,Losada
2018}). More recently, a growing interest is put in describing
logical structures associated to quantum computation
\cite{Holik-QC-2019,Holik-Toffoli-2017,CagliariGroupA,CagliariGroupB}.

In this work, we give an interesting turn to the quantum logical
approach by introducing a relatively recently discovered logical
technique into the foundations of quantum mechanics. As a result of
our construction, we show that a quantum logical entailment arises
by appealing to the quantum non-deterministic semantics. This is a
step forward in the discussion whether there exists a well-behaved
logic associated with the quantum formalism.

The paper is organized as follows. We start by reviewing the
principle of truth functionality in classical logics in Section
\ref{s:TruthFunctionality}. Then, in Section
\ref{s:ElementaryFacts}, we discuss the Born rule and the
Kochen-Specker and Gleason's theorems from a quantum logical
perspective. There, we discuss in which sense the principle of truth
functionality is not valid in quantum mechanics. In Section
\ref{s:NM}, we review the formalism of N-matrix theory and
non-deterministic semantics. In Section \ref{s:KSandQuantum} we show
how this formalism can be used to describe quantum states, as a
particular set of valuations associated with infinite truth tables,
that we first introduced here. We also discuss extensions to
generalized probabilistic models and the detection of probabilities
with finite precision. In Section \ref{s:LogicalConsequence} we
discuss the possibility of developing a logical consequence in
quantum logic. Finally, in Section \ref{s:Conclusions}, we draw our
conclusions.

\section{The Principle of Truth Functionality and Algebra Homomorphisms}\label{s:TruthFunctionality}

In this section, we review the principle of truth functionality, in
connection with the notion of classical semantics and algebra
homomorphisms. The reader with a background in algebra and logic,
may skip this section. We focus on the homomorphisms between a given
algebra and the two-element Boolean algebra
$\mathbf{B}_{2}=\{0,1\}$, endowed with the usual operations (that we
denote by $\tilde{\vee}$, $\tilde{\wedge}$, and $\tilde{\neg}$). The
principle of truth functionality is very important for our work,
because the KS theorem is expressed in terms of classical valuations
satisfying a functional behavior with respect to truth-value
assignments. We also discuss, in the following, how
non-deterministic semantics can be used to deal with the failure of
the TFP, and how they offer the possibility of understanding the
consequences of the KS theorem under a new light.

In this section, we follow the treatment given in references
\cite{Halmos-LogicAsAlgebra,Sagastume}. Let us start with a
definition of {algebra} that is relevant to our logical approach:

\begin{definition}
A type of algebras is a set $\mathcal{F}$ of function symbols, where
each symbol $f\in\mathcal{F}$ has associated a natural number $n$,
its arity.
\end{definition}

\begin{definition}
Given a type $\mathcal{F}$ of algebras, an algebra of type
$\mathcal{F}$ is a pair $\mathcal{A}=\langle A, F\rangle$, where $A$
is a set and $F=(f^{A}_{i})_{i\in I}$ is a family of operations over
$A$, defined in such a way that, to each symbol of $\mathcal{F}$ of
arity $n$, it corresponds an $n$-ary operation $f_{i}^{A}$. The set
$A$ is the universe associated with the algebra. Usually, one speaks
about the algebra by referring to its universe only, in case that
the operations are clearly understood from the context. We also
write $f_{i}$ instead of $f_{i}^{A}$, where no confusion can arise.
\end{definition}

As an example, consider the algebra $\mathbf{B}_{2}$. Its universe
is the set $\{0,1\}$ and its operations are given by $\tilde{\vee}$,
$\tilde{\wedge}$ and $\tilde{\neg}$, with arity $2$, $2$ and $1$,
respectively.

\subsection{Homomorphisms}

A homomorphism between algebras is defined as follows:

\begin{definition}\label{d:Homomorphism}
Let $\mathcal{A}$=$ \langle A,F\rangle$ and $ \mathcal{B}=$ $\langle
B, G \rangle $ be algebras of the same type and let $h: A
\rightarrow B$ be a function. The map $h$ is a homomorphism from a
$\mathcal{A}$ to $\mathcal{B}$, if for any of $n$-ary symbol
$f^{A}\in F$ we have that, for every n-tuple $(a_{1},...,a_{n})$ of
elements in $A$:
    $$ h(f^{A}_{(a_{1},...,a_{n})})= f^{B}_{(h(a_{1}),...,h(a_{n}))}$$
    being $f^{B}$ the operation in $B$ which corresponds to $ f^{A}$ in $A$.
\end{definition}
 If a homomorphism $h$ is a bijection, it is called an
{isomorphism}.

Each propositional language and its well-formed formulas are defined
by the set of connectives, whenever there is a denumerable set of
propositional variables and punctuation symbols (under the
assumption that we have recursive rules for their formation). Thus,
a type is assigned to each language, in the same way as it occurs
for algebras. For more discussion about this, we refer the reader to
\cite{Montano}. In order to illustrate these ideas, let us consider
the {implicational propositional calculus}. The only connective of
this calculus is ``$\rightarrow$'', which is binary. Thus, each
formula which is not a variable is of the form $a\rightarrow b$.
This is a type $ \langle 2\rangle$ language. The classical
propositional calculus with its connectives ``$\neg$'', ``$\vee$'',
``$\wedge$'' and ``$\rightarrow$'', is of the type $\langle
1,2,2,2\rangle$.

We can evaluate the formulas of a language $L$ in algebras of the
same type, proceeding similarly as with homomorphisms between
algebras. Valuations are defined in such a way that each $n$-ary
connective is transformed into a corresponding $n$-ary operation.
Valuations assign to each formula of the language an element of an
algebra, that we may think of as its truth-value. In the classical
case, this algebra is $\mathbf{B}_{2}$, and $0$ and $1$ are
identified as the values ``true'' and ``false'', respectively. In
the classical propositional calculus, this construction gives place
to the well-known truth tables.

At this point, some readers might be interested in understanding
with more detail the link between the type of a language and a type
of algebra. For self completeness, we have included section
\ref{s:Tipos} below, in which we explain these notions with more
detail.

We now give the definition of valuation for a propositional
language:

\begin{definition}\label{d:TruthFunctionality}
Let $L$ be a propositional language whose set of connectives is
$(c_{i})_{i\in I}$ and let $\langle A,G\rangle$ be an algebra where
$G=(g_{i})_{i\in I}$. A (deterministic) {valuation} is a function
$v:L\rightarrow A$, such that for each $n$-ary connective $c$ and
formulas $B_{1},...B_{n}$, it satisfies
$$ v(c(B_{1},...,B_{n}))= g^{c}(v(B_{1}),...,v(B_{n})),$$
being $g^{c}$ the $n$-ary operation in $A$ corresponding to $c$.
\end{definition}

From the above definition of {classical valuation}, it follows that
the value of $g^{c}(v(B_{1}),...,v(B_{n}))$ is {determined} by the
values $v(B_{1})$,....,$v(B_{n})$, that the valuation $v$ assigns to
the propositions $B_{1}$,....,$B_{n}$ out of which the formula
$c(B_{1},...,B_{n})$ is composed. {This is the exact content of the
principle of truth functionality}. A truth functional operator is an
operator whose values are determined by those of its components. All
classical operators are truth-functional. Accordingly, classical
propositional logic is a {truth-functional propositional logic}.

In classical propositional logic the algebra is given by
$\mathbf{B}_{2}$. Thus, for example, the value $v(P\vee Q)$ assigned
by a classical valuation $v$ to the compound proposition $P\vee Q$
is determined solely by the values $v(P)$ and $v(Q)$. In other
words, we have $v(P\vee Q)=v(P)\tilde{\vee}v(Q)$. Similar examples
can be given for $\tilde{\wedge}$ and $\tilde{\neg}$. In Section
\ref{s:ElementaryFacts} we see how these notions can be extended to
the quantum formalism.

\subsubsection{Types of Languages and Homomorphisms between Structures}\label{s:Tipos}

For a detailed treatment of the content of this section, see
\cite{Montano}. Let us start with the definition of~type:

\begin{definition}
A type (or signature) is a set $\tau$ of symbols which has the form:
$$\tau =(\bigcup_{1\leq n}R_{n})\cup (\bigcup_{1 \leq m}F_{m}) \cup C,$$
where $R_{n}$ is a set of relational $n$-arity symbols, $F_{m}$ is a
set of $m$-arity functional symbols and $C$ is a set of symbols for
constants (or any other distinguished symbols of the system), which
are referred as $0$-arity functions.
\end{definition}

The relational and functional symbols acquire meaning only when they
are considered in connection with a semantics or an interpretation.
Thus, they are interpreted as relations, functions and distinguished
elements, respectively, in a given universe of interpretation. It is
necessary to define a language of a given type in order to apply the
definitions of valuation and homomorphism, and then, to relate the
notion of language with that of algebra (or structure in the more
general case).

In this work, we will restrict ourselves to the propositional
calculus only. We include here the definition of first-order
languages, because it can be useful for the development of
quantum-inspired~languages.

\begin{definition}
The symbols for building expressions of a language of type
$\tau$---that we denote by $L_{\tau}$---are the~following:
\begin{enumerate}
\item Individual variables: $v_{0},v_{1},...,v_{n},...$.
\item Auxiliary symbols: left and right parenthesis, and commas.
\item Propositional connectives: $\neg$, $\vee$, $\wedge$, $\rightarrow$,...
\item Equality symbol: $=$.
\item Existential quantifier: $\exists$.
\item The symbols of $\tau$.
\end{enumerate}
\end{definition}

The symbols contained in $1$ to $4$ above are usually referred to as
the {canonical interpretation} and are called {logical symbols}; the
symbols contained in $5$ and $6$ may have a variable interpretation,
and are called non-logical symbols. This is the reason why $\tau$
determines the type of language.

Some remarks are in order. In the quantum formalism, a natural
choice is to consider the orthogonal projections as individual
variables. In classical logic, it is usual to work with only two
connectives, as for example, $\neg$ and $\vee$, and to define the
rest of the connectives as a function of them. This is usual in
proof theory to simplify the object language. The possibility of
doing this simplification depends on the specific properties of the
given language (i.e., connectives are not always inter-definable).
In this work, we will restrict the use to $\vee$, $\neg$ and
$\wedge$. Regarding point $4$ above: languages containing this
symbol are called {languages with an equality}. Regarding $5$: we
will not use quantifiers in this work for the quantum case.

Now we give the standard definitions of terms, expressions and
formulas.

\begin{definition}
Let $\tau$ be a type. An expression of type $\tau$, or a
$\tau$-expression, is a finite sequence of symbols of $L_{\tau}$.
\end{definition}

\begin{definition}
The set of terms of type $\tau$, or $\tau$-terms, is the least set
$X$ of $\tau$-expressions satisfying:
\begin{itemize}
\item $\{v_{i} \,:\, i \geq 0\}\cup C\subseteq X$, where $C\subseteq\tau$, is the set of constants of $\tau$.
\item If $f\in F_{m}\subseteq \tau$, $1\leq m$ and $t_{1},...,t_{m}\in
X$, then $f(t_{1},...,t_{m})\in X$.
\end{itemize}
\end{definition}

\begin{definition}
A $\tau$-formula is atomic if it is an expression of the form: \\
$(t_{1}= t_{2})$ or $P(t_{1},...,t_{n})$, where $t_{1},...,t_{n}$
are $\tau$-terms and $P\in R_{n}\subseteq \tau$.
\end{definition}
\begin{definition}
The set of formulas of type $\tau$ is the least set $X$ of
$\tau$-expressions, such that:
\begin{itemize}
\item $\{\alpha : \alpha \hbox{ } \mbox{is an atomic} \hbox{ } \tau\mbox{-formula}\}\subseteq X$.
\item If $\alpha,\beta\in X$, then $(\neg \alpha)$, $(\alpha\vee \beta)$, $(\alpha \wedge \beta)$ and $(\alpha \rightarrow \beta) \in X$.
\item If $\alpha \in X $ and $v_{i}$ is a variable, then $(\exists
v_{i}\alpha )\in X$.
\end{itemize}
\end{definition}

\begin{definition}
A $\tau$-interpretation (or $\tau$-structure) for a language $L$ is
a pair $\mathcal{U}=\langle A,I\rangle$, where:

\begin{itemize}
\item $A\not = \emptyset$.
\item $I:\tau \rightarrow A \cup \{f: A^{m}\rightarrow A , 1\leq m\} \cup (\bigcup \{P(A^{n}):1\leq n\})$.
\end{itemize}

\noindent $P(A^{n})$ denotes the power set of $A^{n}$, and such that for any $x\in \tau $:\\
If $x\in R_{n}$, then $I(x)= x^{\mathcal{U}}\subseteq A^{n}$.\\
If $x \in F_{m}$, then $I(x)=x^{\mathcal{U}}: A^{m}\rightarrow A$.\\
If $x\in C$, then $I(x)=x^{\mathcal{U}}\in A$.\\
I is the interpretation function of $\mathcal{U}$ over the universe
$A$.

\end{definition}

Usually, if $\tau= \{x_{1},...,x_{n}\}$, we denote the structure
$\mathcal{U}=\langle A,I\rangle$ as follows:
$$\mathcal{U}=\langle A, x^{\mathcal{U}}_{1},...,x^{\mathcal{U}}_{n}\rangle$$

In our case, where the language is formed by projection operators
acting on a Hilbert space endowed with their respective connectives,
the structure $\mathcal{U}$ will have a universe $A$ and
interpretations for the connectives, which are symbols of functions.
In such a structure there will be no interpreted relation symbols.
For this reason, the structure associated with our language is an
algebra (and not a more general structure), and we can relate it
with algebras of the same type as we did in defining valuations and
homomorphisms for the language of our type.

\begin{definition}
Let $\mathcal{U}=\langle A,I \rangle$ and $\mathcal{V}=\langle
B,J\rangle$ be two $\tau$-structures. Then, $h$ is a homomorphism
from $\mathcal{U}$ to $\mathcal{V}$ if, and only if:
\begin{itemize}
\item $h$ is a function from $A$ to $B$; $h:A\rightarrow B$.
\item For each relational n-ary symbol $r_{n}$ in $\tau$ and each $a_{1},...,a_{n}\in A$,
$$ \langle a_{1},..., a_{n}\rangle \in r_{n}^{\mathcal{U}} \hbox{ iff } \langle h(a_{1}),...,h(a_{n})\rangle \in r_{n}^{\mathcal{V}}.$$
\item For each functional n-ary symbol $f_{n}$ in $\tau$ and each $a_{1},...,a_{n}\in A$, $$ h(f_{n}^{\mathcal{U}}(a_{1},...,a_{n}))=f^{\mathcal{V}}_{n}(h(a_{1}),...,h(a_{n})).$$
\item For each $c\in \tau , h(c^{\mathcal{U}})=c^{\mathcal{V}}.$
\end{itemize}
\end{definition}

 It can be seen how this definition for homomorphisms
between structures generalizes the one that we gave for a
homomorphism between algebras. This is a consequence of the fact
that we are dealing with a language of type $\tau$ that only has
function symbols (which are the connectives of our language).

\section{Quantum States and the Gleason and Kochen--Specker Theorems}\label{s:ElementaryFacts}

In this section, we review two of the most important results in the
foundations of quantum mechanics: Gleason
\cite{Gleason,Gleason-Dvurechenski-2009} and Kochen--Specker
\cite{KST,Svozil-KS,Smith-KS} theorems. The fundamental properties
underlying these theorems will play a key role in the rest of the
paper. We discuss truth functionality in the quantum domain.

\subsection{Quantum Probabilities and Gleason'S Theorem}

An elementary experiment associated with a quantum system is given
by a yes-no test, i.e., a test in which we get the answer ``yes'' or
the answer ``no'' \cite{piron}. As is well known, elementary tests
associated with quantum systems are represented by orthogonal
projections acting on a separable Hilbert space $\mathcal{H}$. An
operator $P\in\mathcal{B}(\mathcal{H})$ is said to be an orthogonal
projection if it satisfies
\begin{equation}
P^{2}=P \qquad \text{and} \qquad P=P^{\ast}.
\end{equation}

Let $\mathcal{P}(\mathcal{H})$ denote the set of all orthogonal
projections acting on $\mathcal{H}$. Denote by
$\mathcal{B}(\mathcal{H})$ the set of bounded operators acting on
$\mathcal{H}$. Projectors and closed subspaces can be put in a one
to one correspondence, by assigning to each orthogonal projection
its image. Thus, they can be considered as interchangeable notions
(and we will use them interchangeably in the following). The set
$\mathcal{P}(\mathcal{H})$ can be endowed with an orthocomplemented
lattice structure $\mathcal{L}(\mathcal{H})=
\langle\mathcal{P}(\mathcal{H}),\ \wedge,\ \vee,\ \neg,\
\mathbf{0},\ \mathbf{1}\rangle$, where $P\wedge Q$ is the orthogonal
projection associated to intersection of the ranges of $P$ and $Q$,
$P\vee Q$ is the orthogonal projection associated with the closure
of the direct sum of the ranges of $P$ and $Q$, $\mathbf{0}$ is the
null operator (the bottom element of the lattice), $\mathbf{1}$ is
the identity operator (the top element), and $\neg(P)$ is the
orthogonal projection associated with the orthogonal complement of
the range of $P$ \cite{mikloredeilibro}. It is important to remark
that the symbol ``$\mathcal{L}(\mathcal{H})$'' denotes the
collection of orthogonal projections of $\mathcal{H}$ and it should
not be confused with the set of linear operators acting on it. This
lattice (which is equivalent to that of closed subspaces of
$\mathcal{H}$) was discovered by Birkhoff and von Neumann, who
coined the term {Quantum Logic} \cite{BvN}. The main characteristic
of this quantum structure is that it is a non-Boolean lattice. It is
always modular in the finite-dimensional case and never modular in
the infinite one \cite{mikloredeilibro}.

Any observable quantity can be represented by a self-adjoint
operator. For every self-adjoint operator $A$, if the system is
prepared in state $\rho$, its mean value is given by the formula:
\begin{equation}
\langle A\rangle=\mbox{tr}(\rho A).
\end{equation}

Due to the spectral theorem, self-adjoint operators are in one to
one correspondence with projective valued measures (PVM) \cite{vN}.
Let $\mathcal{B}$ be the Borel set of the real line. Given a
self-adjoint operator $A$, its {projection-valued measure} $M_{A}$
is a map \cite{mikloredeilibro}:
\begin{equation}
M_{A}: \mathcal{B}\mapsto \mathcal{P}(\mathcal{H}),
\end{equation}
such that
\begin{enumerate}

\item $M_{A}(\emptyset)=\mathbf{0}$

\item $M_{A}(\mathbb{R})=\mathbf{1}$

\item $M_{A}(\cup_{j}(B_{j}))=\sum_{j}M_{A}(B_{j})$,\,\, \mbox{for any mutually disjoint
family} ${B_{j}}$

\item $M_{A}(B^{c})=\mathbf{1}-M_{A}(B)=(M_{A}(B))^{\bot}$.

\end{enumerate}

In quantum mechanics, states can be considered as functions that
assign probabilities to the elements of $\mathcal{L}(\mathcal{H})$.
A state on a quantum system is represented by a function:
\begin{equation}\label{e:nonkolmogorov}
\mu:\mathcal{L}(\mathcal{H})\longrightarrow [0,1]
\end{equation}
satisfying:
\begin{enumerate}

\item $\mu(\textbf{0})=0$.

\item $\mu(P^{\bot})=1-\mu(P)$

\item For any pairwise orthogonal and denumerable family $\{P_{j}\}_{j\in\mathbb{N}}$,
$\mu(\bigvee_{j}P_{j})=\sum_{j}\mu(P_{j})$.
\end{enumerate}

{Gleason's theorem \cite{Gleason,Gleason-Dvurechenski-2009} is
equivalent to the assertion that whenever $dim(\mathcal{H})\geq 3$,
the set $\mathcal{C}(\mathcal{L}(\mathcal{H}))$ of all measures of
the form \eqref{e:nonkolmogorov} can be put in a one to one
correspondence with the set $\mathcal{S}(\mathcal{H})$ of all
positive and trace-class operators of trace one acting in
$\mathcal{H}$}. The correspondence is such that for every measure
$\mu$ satisfying the above axioms, there exists a density operator
$\rho_{\mu}\in\mathcal{S}(\mathcal{H})$ such that for every
orthogonal projector $P$ representing an elementary test, we have:
\begin{equation}\label{e:bornrule}
\mu(P)=\mbox{tr}(\rho_{\mu}P).
\end{equation}

It will be important for us to recall how probabilities are defined
in a classical setting. Thus, given a set $\Omega$, let us consider
a $\sigma$-algebra $\Sigma\subseteq\mathcal{P}(\Omega)$ of subsets.
Then, a {Kolmogorovian probability measure} will be given by a
function
\begin{equation}\label{e:kolmogorovian}
\mu:\Sigma\rightarrow[0,1]
\end{equation}
satisfying
\begin{enumerate}

\item $\mu(\emptyset)=0$

\item $\mu(A^{c})=1-\mu(A), \,\, \mbox{where}\, (\ldots)^{c}\, \mbox{denotes set theoretical
complement}$

\item for any
pairwise disjoint and denumerable family
$\{A_{i}\}_{i\in\mathbb{N}}$,
$\mu(\bigcup_{i}A_{i})=\sum_{i}\mu(A_{i})$.

\end{enumerate}

The above conditions \eqref{e:kolmogorovian} are known as
Kolmogorov's axioms \cite{KolmogorovProbability} and are very useful
for theoretical purposes. The main difference between classical and
quantum probabilities---pass their similitude in shape---comes from
the fact that the $\sigma$-algebra $\Sigma$ appearing in
\eqref{e:kolmogorovian} is Boolean, while $\mathcal{L}(\mathcal{H})$
is not. This is the reason why quantum probabilities are also called
non-Kolmogorovian (or non-Boolean) probability measures (for more
discussion on this subject, see for example
\cite{Gudder-StatisticalMethods,Redei-Summers2006,Holik-AOP-2014,ReviewNOVA};
for the connection between quantum probabilities and quantum
information theory, see \cite{Holik-Entropy-2015,HolikQIC-2016}).

The lattice $\mathcal{L}(\mathcal{H})$ of closed subspaces (or
equivalently, orthogonal projections) of a separable Hilbert space
and the Boolean algebras (such as those appearing in
\ref{e:kolmogorovian}), are all examples of the more general family
of {orthomodular lattices} \cite{kalm83}. A lattice $\mathcal{L}$ is
said to be {orthomodular} (or {weak modular}) if it is
orthocomplemented and, whenever $x\leq y$, then $y=x\vee (y \wedge
x^{\bot})$. It turns out that the orthogonal projections associated
to a von Neumann algebra form an orthomodular lattice
\cite{mikloredeilibro}. Von Neumann algebras play a key role in the
rigorous treatment of quantum systems with infinitely many degrees
of freedom \cite{Redei-Summers2006}. Boolean algebras---such as
those appearing in \ref{e:kolmogorovian}---are particular cases of
orthomodular lattices. Thus, we can conceive the state of a general
physical system described by a measure over an arbitrary complete
orthomodular lattice $\mathcal{L}$ as follows:
\begin{equation}\label{e:generalizedstates}
\mu:\mathcal{L}\rightarrow[0,1]
\end{equation}
which satisfies
\begin{enumerate}

\item $\mu(\mathbf{0})=0$

\item $\mu(a^{\bot})=\mathbf{1}-\mu(a), \,\, \mbox{where}\, (\ldots)^{\bot}\, \mbox{denotes
the orthocomplement}$

\item for any pairwise orthogonal and denumerable family of elements $\{a_{i}\}_{i\in I}$, $\mu(\bigvee_{i}a_{i})=\sum_{i}\mu(a_{i})$.

\end{enumerate}

When $\mathcal{L}$ is Boolean, we obtain a classical probabilistic
model. The lattice $\mathcal{L}(\mathcal{H})$ is a particular case
among a vast family of alternative models of physical systems.

As a consequence of the orthomodular law, it trivially follows that
for every $p,q\in\mathcal{L}$ and every state $\mu$, we have
$p\wedge q \leq p \Rightarrow \mu(p\wedge q)\leq \mu(p)$ and $p\leq
p\vee q \Rightarrow \mu(p)\leq \mu(p\vee q)$. These simple
inequalities will be used in sections \ref{s:KSandQuantum} in order
to build N-matrices for quantum and generalized probabilistic
models.

To finish this section, let us recall an important property of
orthomodular lattices that are useful to keep in mind in the
following. Any orthomodular lattice---even if it is
non-Boolean---possesses Boolean subalgebras \cite{kalm83}. A state
defined by Equations \eqref{e:generalizedstates}, when restricted to
a {maximal} Boolean subalgebra, defines a Kolmogorovian probability
measure satisfying \eqref{e:kolmogorovian}. Moreover, in the quantum
formalism, a maximal observable $A$ defines a maximal Boolean
subalgebra $\mathcal{B}_{A}\subseteq\mathcal{L}(\mathcal{H})$, which
is given by the range of $M_{A}$~\cite{piron}.

\subsection{The Kochen--Specker Theorem and the Failure of Truth Functionality in Quantum Mechanics}

The Kochen--Specker theorem is one of the cornerstones in the
foundations of quantum mechanics literature \cite{KST}. Kochen and
Specker were looking for a description of quantum mechanics in terms
of hidden variables, taking as a model the relationship between
classical statistical mechanics and thermodynamics. It turns out
that this hidden variable theory cannot exist for the quantum case.
In~order to understand the mathematical structures behind the KS
theorem, let us define first:

\begin{definition}\label{d:KeyProperty}
A classically truth-valued function, is a function
$v:\mathcal{L}(\mathcal{H})\longrightarrow\{0,1\}$ having the
property that $\sum_{i}v(P_{i})=1$ for any family
$\{P_{i}\}_{i\in\mathbb{N}}$ of one dimensional orthogonal elements
of $\mathcal{L}(\mathcal{H})$ satisfying $\sum_{i}P_{i}=\mathbf{1}$.
\end{definition}

The no-go theorem for the hidden variables described in the KS paper
is equivalent to the following statement \cite{KST}.

\begin{prop}\label{prop:KStheorem}
There exists no classically truth-valued function.
\end{prop}

In order to understand what the KS theorem says from a logical point
of view, let us see how the notion of truth functionality can be
conceived in the quantum domain. Let us identify the quantum
language with that of the lattice structure
$\mathcal{L}(\mathcal{H})=\langle\mathcal{P}(\mathcal{H}),\vee,\wedge,\neg,\mathbf{0},\mathbf{1}\rangle$.
It is clear that we can form, recursively, new propositions out of
any given set of propositions in the usual way (i.e., consider all
possible finite expressions such as $(P\vee Q)\wedge R$, $(\neg
P\wedge Q)$, and so on). If we want to recreate the notion of
classical valuation in quantum theory, there should exist a set of
functions $v:\mathcal{L}(\mathcal{H})\longrightarrow\{0,1\}$
assigning truth values to all possible elementary propositions. In
this way, we would have that, given a quantum system prepared in a
particular state, all of its properties should satisfy being true or
false, and there should be no other possibility. From the physical
point of view, there are several extra conditions to be imposed on
these classical valuations. For example, if no other restrictions
are imposed, one can have the valuation $v(P)=0$ for all $P$, which
is nonsensical because it would imply that all outcomes will have
zero probability of occurrence in any experiment. Which restrictions
are to be imposed? We need to define a set of conditions in order to
discard the non-physical valuations.

According to the probabilistic rules of quantum mechanics (and any
physically consistent theory), if proposition $P$ is true (i.e., if
$v(P)=1$), then, any other proposition $Q$ satisfying $Q\leq
P^{\bot}$ should be false ($v(Q)=0$). This follows directly from the
definition of quantum state: if $P$ is true, its probability of
occurrence is equal to one, and the probability of occurrence of any
orthogonal property will be automatically zero (this follows
directly from Equation \eqref{e:nonkolmogorov}; a similar conclusion
holds in generalized models by using Equation
\eqref{e:generalizedstates}).

Another condition to be imposed in order to obtain properly
classical valuations, should be that of being an algebra
homomorphism between $\mathcal{L}(\mathcal{H})$ and the two-element
Boolean algebra $\mathbf{B}_{2}=\{0,1\}$ (see definition
\ref{d:Homomorphism}). Thus, for every $P$ and every family
$\{P_{i}\}_{i=1}^{n}$, {in analogy with the principle of truth
functionality given by definition \ref{d:TruthFunctionality}}, we
should have:
\begin{equation}\label{e:QTruthF}
v(\bigvee_{i}P_{i})=\tilde{\bigvee}_{i}v(P_{i})
\end{equation}
\begin{equation}\label{e:QTruthFwedge}
v(\bigwedge_{i}P_{i})=\tilde{\bigwedge}_{i}v(P_{i})
\end{equation}
\begin{equation}\label{e:QNeg}
v(\neg P)=\tilde{\neg}v(P)=1-v(P)
\end{equation}

We call {classical admissible valuations} (and denote it by $CAV$)
the set of two-valued functions satisfying Equations
\eqref{e:QTruthF}--\eqref{e:QNeg}. Notice that, if the set of
admissible valuations is assumed to be $CAV$, truth-functionality is
automatically satisfied.

The condition represented by Equation \eqref{e:QNeg} implies that
for a denumerable orthonormal and complete set of projections
$P_{i}$ (i.e., $\sum_{i}P_{i}=\mathbf{1}$, $P_{i}P_{j}=\mathbf{0}$
and $\mbox{dim}(P_{i})=1$), if $v(P_{i_{0}})=1$ for some $i_{0}$,
then, $v(P_{j})=0$, for all $j\neq i_{0}$ (this follows from the
fact that for $j\neq i_{0}$, $P_{j}\leq \mathbf{1}-P_{i_{0}}$).
Given that $\sum_{i}P_{i}=\mathbf{1}$, we {must} have
$v(P_{i_{0}})=1$, for some $i_{0}$ (this follows from
$v(\mathbf{1})=1$ and Equation \eqref{e:QTruthF}). This is a very
reasonable physical property. It implies that, if an experiment is
performed on the system (remember that any denumerable orthonormal
and complete set of one-dimensional projections defines an
experiment), we obtain that one, and only one projection operator is
assigned the value $1$, while all other outcomes (which define
exclusive events), are assigned the value $0$. Notice that this is
the condition of KS (i.e., the condition appearing in definition
\ref{d:KeyProperty}). It is important to remark that an experiment
in which all propositions are assigned the value false, or an
experiment in which more than one exclusive alternatives are
assigned the value true, are physically nonsensical. In the latter
case, we would obtain that two mutually exclusive alternatives would
occur with complete certainty. In the former, we would reach a
situation in which all outcomes of an experiment is false. From the
perspective of a classical ontology, these alternatives should be
discarded.

We have seen above that any admissible valuation in $CAV$ should
satisfy definition \ref{d:KeyProperty}. The existence of such
functions is {strictly forbidden} by the KS theorem. It follows that
two-valued functions satisfying both, physically reasonable
requirements and truth functionality, cannot exist in the quantum
domain. This will be naturally true for arbitrary probabilistic
models, provided that their propositional structures satisfy the KS
theorem (and this is quite true for a huge family of
models~\cite{Doring-KSVNA,Svozil-KS,Smith-KS}).  Notice that the KS
argument works for Hilbert spaces of dimension greater than or equal
to $3$. A simple check also shows that no element in $CAV$ exists
for a qubit system (i.e., a two-dimensional quantum model). This is
so because conditions \eqref{e:QTruthF}--\eqref{e:QNeg} are even
stronger than those involved in definition \ref{d:KeyProperty}.

What about weaker conditions? Perhaps, if we give up some physical
considerations and focus only on pure mathematics, we might find a
set of admissible functions with regard to which the connectives
behave truth functionally. In \cite{Friedman-Glymour}, Friedman, and
Glymour study which are the most reasonable conditions to be imposed
on the set of admissible valuations. After discarding non-physical
possibilities, they end up with the conditions:

$v:\mathcal{P}(\mathcal{H})\longrightarrow \{0,1\}$ is an admissible
valuation iff
\begin{itemize}
\item for every $P$, $v(P)=1$ iff $v(\neg P)=0$
\item for every pair $P,Q\in \mathcal{P}(\mathcal{H})$, if $v(P)=1$ and $P\leq
Q$, then $v(Q)=1$.
\end{itemize}

Let us call $S3$ the set of valuations satisfying the above-defined
conditions. In \cite{Friedman-Glymour}, $S3$ is used to denote the
list of conditions, but this small difference in notation should not
lead to confusion here. Notice first that the conditions that define
$CAV$ (i.e., Equations \eqref{e:QTruthF}--\eqref{e:QNeg}), imply
those that define $S3$, and thus are stronger. In order to study
whether $S3$ is a non-empty set or not, Friedman and Glymour make
further distinctions. First, they consider the {normal} admissible
valuations, which are those satisfying $S3$, plus the condition that
they assign the truth value $1$ to at least one one-dimensional
subspace. Again, a valuation not satisfying this minimal requirement
cannot belong to the realm of a classical ontology. Let us denote by
$NS3$ the set of normal valuations. Clearly, $NS3\subseteq S3$.
Friedman and Glymour show---by construction---that $NS3$ is
non-empty. Later, they discuss whether it is possible that the
valuations in $S3$ satisfy the minimal physical requirement of
realism that every observable have a precise value. For this to
happen, it should be the case that for every orthogonal basis,
exactly one vector receives the value true and the remainder receive
the value false. Let us call $RS3$ the set of normal admissible
functions satisfying this physical requirement. Notice that
$RS3\subseteq NS3 \subseteq S3$. Again, Friedman and Glymour remark
that $RS3$ is an empty set, due to the KS theorem. By the above
discussion, we also have that the conditions defining $CAV$
(Equations \eqref{e:QTruthF}--\eqref{e:QNeg}) imply those of $RS3$,
but the converse is not true.

What is left? We are only left then with the functions in $NS3$,
that satisfy the undesirable condition that some observables do not
have one (and only one) true projection operator. The whole program
of having a reasonable classical valuation satisfying reasonable
physical conditions is lost (due to the KS theorem). Even so, if we
only restrict to the purely mathematical entities in $NS3$,
G.~Hellman showed in \cite{Hellman-TF} that the quantum logical
connectives do not behave truth-functionally with regard to those
admissible functions. One of the admitted conclusions of Hellman's
work is that his theorem belongs to pure mathematics, with no
connection to quantum mechanics.

All roads lead to the same conclusion: it is not possible to define
classical truth-functional valuations satisfying reasonable physical
requirements (in the sense of $CAV$ and $RS3$), given that the KS
theorem blocks their existence. If one gives up the minimal physical
requirements of a classical ontology (using for example, NS3), these
functions will not be truth-functional either.

Some important remarks are at stake. The above discussion is related
to a well-known fact: it is possible to define {local} classical
valuations for maximal Boolean subalgebras of
$\mathcal{L}(\mathcal{H})$, but the Kochen-Specker theorem forbids
the existence of global ones (see sections II and III of \cite{KST};
for more discussion on the subject, see
\cite{Isham-Topos-1998,domenech-Freytes-2005,Domenech-2008,Bub-Clifton,Redhead-Book}).
Another related fact is that of the non-existence of states whose
range is equal to the set $\{0,1\}$ (also known as {dispersion-free
states} in the foundations of quantum physics literature). There are
several proofs of the non-existence of dispersion free-states, and
the KS and Gleason's theorems can be considered among them. As far
as we know, the oldest one is due to J. von Neumann \cite{vN}, and
it is very important to mention here the works of J. Bell on the
subject~\cite{Bell-64,Bell-66}. All these works involve different
assumptions and mathematical techniques. In this work, we have put
the focus on the works of Kochen and Specker, because their approach
is the one best fitting to the algebraic structures that we use in
order to build the N-matrices for quantum and generalized
probabilistic models.

Indeed, if one looks carefully at the KS theorem, it is possible to
recognize that a very particular form of the principle of truth
functionality fails in the quantum realm. Namely, that there is no
classical truth-value assignment $v$ satisfying the functionality
condition given in definition \ref{d:KeyProperty} (or equivalently,
the conditions \eqref{e:QTruthF}--\eqref{e:QNeg}). Indeed, in the KS
paper \cite{KST} (see also \cite{Isham-Topos-1998}), it is proved
that this condition underlies the failure of a more general
property. In order to illustrate the idea, suppose that the
observable represented by the self-adjoint operator $A$ has
associated the real value of $a$. Then, the observable represented
by $A^{2}$, should have assigned the value $a^{2}$. In this sense,
observables are not all independent, and neither are the values
assigned to them. This gives us a clue for understanding why
truth-functionality is not valid in the quantum domain. Following
the spirit of the KS paper, let us~define:

\begin{definition}\label{d:TruthFunctionalityOperators}
Let $\mathcal{A}(\mathcal{H})$ be the set of all self-adjoint
operators acting on a separable Hilbert space $\mathcal{H}$. A
function $f:\mathcal{A}(\mathcal{H})\longrightarrow\mathbb{R}$
satisfies {truth functionality} if, for any Borel function
$g:\mathcal{A}(\mathcal{H})\longrightarrow\mathcal{A}(\mathcal{H})$,
if $g(A)$ is the result of applying the function $g$ to a
self-adjoint operator $A$ in the usual way, and $f_{X}$ is the
result of applying $f$ to an arbitrary self-adjoint operator $X$,
the condition $f_{g(A)}=g(f_{A})$ is satisfied.
\end{definition}

A function such as $f$ in definition
\ref{d:TruthFunctionalityOperators} can be called a {prediction
function} (see section II in \cite{KST}) because it assigns a given
value to each quantum observable. Condition
\begin{equation}\label{e:FunctionalityCondition}
f_{g(A)}=g(f_{A})\,\,,\,\,\mbox{for every Borel function $g$},
\end{equation}
imposes a strong condition on $f$. Let us see how this works (see
also sections I and II in \cite{KST}). As is well known, two quantum
mechanical observables represented by self-adjoint operators $A$ and
$B$, respectively, are compatible, if and only if, there exist Borel
functions $g_{1}$ and $g_{2}$, and a self adjoint operator $C$, such
that $A=g_{1}(C)$ and $B=g_{2}(C)$. Thus, whenever $A$ and $B$ are
compatible, using the functionality condition
\eqref{d:TruthFunctionalityOperators} we have
$f_{AB}=f_{g_{1}(C)g_{2}(C)}=f_{(g_{1}g_{2})(C)}=(g_{1}g_{2})(f_{C})=g_{1}(f_{C})g_{2}(f_{C})=f_{g_{1}(C)}f_{g_{2}(C)}=f_{A}f_{B}$.
If $\alpha$ and $\beta$ are real numbers, we also have, for
compatible $A$ and $B$, that $f_{\alpha A+\beta B}=f_{\alpha
g_{1}(C)+\beta g_{2}(C)}=f_{(\alpha g_{1}+\beta g_{2})(C)}=(\alpha
g_{1}+\beta g_{2})(f_C)=\alpha g_{1}(f_C)+\beta g_{2}(f_C)=\alpha
f_{g_{1}(C)}+\beta f_{g_{2}(C)}=\alpha f_{ A}+\beta f_{B}$. It
follows that $f$ is a partial Boolean algebra homomorphism.
Furthermore, if $P^{2}=P$ and $P^{\dag}=P$ (i.e., if $P$ is an
orthogonal projection), we have
$f_{P}=f_{P^{2}}=f_{P}f_{P}=f_{P}^{2}$, and then, $f_{P}=0$ or
$f_{P}=1$. It should be clear that a function with these properties
cannot exist because it goes against
Proposition~\ref{prop:KStheorem} (see also definition
\ref{d:KeyProperty}). Thus, we see that the failure of the above
described {sui generis} version of truth functionality (the one
contained in definition \eqref{d:TruthFunctionalityOperators} and
Equation \eqref{e:FunctionalityCondition}), is one of the key
features of quantum mechanics. This is true for more general
probabilistic models provided that their propositional structures
admit no global classical valuations to $\mathbf{B}_{2}$ (i.e.,
provided that they do not admit valuations satisfying Equations
\eqref{e:QTruthF}--\eqref{e:QNeg}).

In the following sections, we exploit the failure of truth
functionality in quantum mechanics and import into physics the
solution given by logicians to its equivalent failure in logical
systems: we will connect the semantics of N-matrix approach in logic
with the formalism of quantum mechanics.

\section{Non-Deterministic Semantics}\label{s:NM}

Non-deterministic multi-valued matrices (N-matrices) are a fruitful
and quickly expanding field of research introduced in
\cite{AvronLev,Avron-Lev-2001b,Avron-Konikowska-2005} (see also
\cite{Batens-1999,Crawford-1998}). Since then it has been rapidly
developing towards a foundational logical theory and has found
numerous applications \cite{Avron-Zamansky}. The novelty of
N-matrices is in extending the usual many-valued algebraic semantics
of logical systems by importing the idea of non-deterministic
computations, and allowing the truth-value of a formula to be chosen
non-deterministically out of a given set of options. N-matrices have
proved to be a powerful tool, the use of which preserves all the
advantages of ordinary many-valued matrices, but is applicable to a
much wider range of logic. Indeed, there are many useful
(propositional) non-classical logics, which have no finite
multi-valued characteristic matrices, but do have finite N-matrices,
and thus are~decidable.

\subsection{Deterministic Matrices}

Here we follow the presentation of the subject given in
\cite{Avron-Zamansky}. In what follows, $L$ is a propositional
language and $Frm_{L}$ is its set of well formed formulas. The
metavariables $\varphi$, $\psi$,..., range over $L$-formulas, and
$\Gamma$, $\Delta$,..., over sets of $L$-formulas. The standard
general method for defining propositional logics is by using
(possibly many-valued) deterministic matrices:

\begin{definition}
A matrix for L is a tuple
$$P = \langle V;D;O\rangle ,$$
where
\begin{itemize}
\item $V$ is a non-empty set of truth-values.
\item $D$ (designated truth-values) is a non-empty proper subset of V.
\item For every n-ary connective $\diamondsuit$  of $L$, $O$ includes a
corresponding function $\widetilde{\diamondsuit}: V^{n}\rightarrow
V$
\end{itemize}

A partial valuation in $P$ is a function $v$ to $V$ from some subset
$\mathcal{W}\subseteq Frm_{L}$ which is closed under subformulas,
such that for each n-ary connective $\diamondsuit$  of $L$, the
following holds for all $\psi_{1},...,\psi_{n}\in \mathcal{W}:$

\begin{equation}
v(\diamondsuit (\psi_{1},...,\psi_{n}))=
\widetilde{\diamondsuit}(v(\psi_{1}),...,v(\psi_{n}))
\end{equation}

\end{definition}

\subsection{Non-Deterministic Matrices (N-Matrices)}

Now we turn into the non-deterministic case. The main difference is
that, alike deterministic matrices, the non-deterministic ones,
given the inputs in the truth table, assign a set of possible values
instead of a single one.

\begin{definition}
A non-deterministic matrix (N-matrix) for $L$ is a tuple $M=\langle
V,D,O\rangle$, where:
\begin{itemize}
    \item $V$ is a non-empty set of truth-values.
     \item $D\in \mathcal{P}(V)$ (designated truth-values) is a non-empty proper subset of $V$.
    \item For every n-ary connective $\diamondsuit$ of $L$, $O$ includes a corresponding function $$\widetilde{\diamondsuit}:V^{n}\rightarrow \mathcal{P}(V)\setminus\{\emptyset\}$$
\end{itemize}
\end{definition}

\begin{definition}\quad

\begin{enumerate}
\item A partial dynamic valuation in $M$ (or an $M$-legal partial dynamic
valuation), is a function $v$ from some closed under subformulas
subset $\mathcal{W}\subseteq Frm_{L}$ to $V$, such that for each
n-ary connective $\diamondsuit$ of $L$, the following holds for all
$\psi_{1},...,\psi_{n} \in \mathcal{W}$:

$$v(\diamondsuit(\psi_{1},...,\psi_{n}))\in
\widetilde{\diamondsuit}(v(\psi_{1}),...,v(\psi_{n})).$$

A partial valuation in $M$ is called a valuation if its domain is
$Frm_{L}$.

\item A (partial) static valuation in $M$ (or an $M$-legal (partial)
static valuation), is a (partial) dynamic valuation (defined in some
$\mathcal{W}\subseteq Frm_{L}$) which satisfies also the following
composability (or functionality) principle: for each n-ary
connective $\diamondsuit$ of $L$ and for every
$\psi_{1},...,\psi_{n}, \varphi_{1},...,\varphi_{n} \in
\mathcal{W}$, if $v(\psi_{i})=v(\varphi_{i})\,\,(i=1,...,n)$, then
$$v(\diamondsuit(\psi_{1},...,\psi_{n}))=v(\diamondsuit (\varphi_{1},...,\varphi_{n})).$$
\end{enumerate}
\end{definition}

 It is important to remark that ordinary (deterministic)
matrices correspond to the case when each $\widetilde{\diamondsuit}:
V^{n}\rightarrow V $ is a function taking singleton values only. In
this case there is no difference between static and dynamic
valuations, and we have full determinism.

To understand the difference between ordinary matrices and
N-matrices, recall that in the deterministic case, the truth-value
assigned by a valuation $v$ to a complex formula is defined as
follows:
$v(\diamondsuit(\psi_{1},...,\psi_{n}))=\widetilde{\diamondsuit}(v(\psi_{1}),...,v(\psi_{n}))
$. Thus the truth-value assigned to
$\diamondsuit(\psi_{1},...,\psi_{n})$ is uniquely determined by the
truth-values of its subformulas: $v(\psi_{1}),...,v(\psi_{n})$.
This, however, is not the case for N-matrices: in general, the
truth-values of $\psi_{1},...,\psi_{n}$, do not uniquely determine
the truth-value assigned to $\diamondsuit(\psi_{1},...,\psi_{n})$
because different valuations having the same truth-values for
$\psi_{1},...,\psi_{n}$ can assign different elements of the set of
options $\widetilde{\diamondsuit}(v(\psi_{1}),...,v(\psi_{n}))$ to
$\diamondsuit(\psi_{1},...,\psi_{n})$. Therefore the
non-deterministic semantics is non-truth-functional, as opposed to
the deterministic one. Notice also that the indeterminism that
appears in the context of N-matrices is defined in terms of the
behavior of valuations, and is related to the failure of
truth-functionality. This notion should not be confused with the
non-deterministic character of quantum phenomena. Whether these
notions can be connected or not, will be the subject of future work.
In Table \ref{tab:DvsND}, we sketch the differences between
deterministic and non-deterministic matrices. Now, we review the
standard definitions of logical consequence \cite{Avron-Zamansky}.

\begin{table}[h!]
  \begin{center}
    \caption{Deterministic vs. non-deterministic matrices.}
    \label{tab:DvsND}
    \begin{tabular}{|l|c|r|} 
  \hline
       & \textbf{Deterministic Matrices} & \textbf{N-Matrices} \\
        \hline
      Truth values set & V & V \\
        \hline
      Designated values set & $D\subset V$ & $D\subset V$\\
        \hline
       Connectives $\diamondsuit$ & $\widetilde{\diamondsuit}:V^{n}\rightarrow V$ & $\widetilde{\diamondsuit}:V^{n}\rightarrow \mathcal{P}(V)\setminus\{\emptyset\}$\\
         \hline
      Valuations & Non-dynamic & Possibly dynamic and possibly non-static \\
        \hline
      Truth-Functional & Yes & Not necessarily \\
        \hline    \end{tabular}
  \end{center}
\end{table}

\begin{definition}\quad

\begin{enumerate}
\item A (partial) valuation $v$ in $M$ satisfies a formula $\psi$
$(v\models \psi$) if ($v(\psi)$ is defined and) $v(\psi)\in D$. It
is a model of $\Gamma$ ($v\models \Gamma$) if it satisfies every
formula in $\Gamma$.
\item We say that $\psi$ is dynamically
(statically) valid in $M$, in symbols $\models^{d}_{M}\psi$
($\models^{s}_{M}\psi$), if $v\models\psi$ for each dynamic (static)
valuation $v$ in $M$.
\item the dynamic (static) consequence relation
induced by $M$ is defined as follows: $\Gamma\vdash^{d}_{M}\Delta$
($\Gamma\vdash^{s}_{M}\Delta$) if every dynamic (static) model $v$
in $M$ of $\Gamma$ satisfies some $\psi\in\Delta$.
\end{enumerate}
\end{definition}

 Obviously, the static consequence relation includes the
dynamic one, i.e., $\vdash^{d}_{M}\subseteq\vdash^{s}_{M}$. For
ordinary matrices, we have $\vdash^{s}_{M}=\vdash^{d}_{M}$.

\begin{theo}
Let M be a two-valued N-matrix which has at least one proper
non-deterministic operation. Then there is no finite family of
finite ordinary matrices F such that $\vdash^{d}_{M}\psi \hbox{ iff
} \vdash_F\psi$.
\end{theo}

\begin{theo}
For every (finite) N-matrix M, there is a (finite) family of
ordinary matrices $F$ such that  $\vdash^{s}_{M}=\vdash_{F}$.
\end{theo}

Thus, only the expressive power of the dynamic semantics based on
N-matrices is stronger than that of ordinary matrices.

\section{N-Matrices for Probabilistic Theories}\label{s:KSandQuantum}

As we have seen in Section \ref{s:ElementaryFacts}, the
Kochen-Specker theorem forbids the existence of a homomorphism
(i.e., valuations satisfying Equations
\eqref{e:QTruthF}--\eqref{e:QNeg}) from the lattice of quantum
propositions to the two-valued algebra $\mathbf{B}_{2}$. There, we
showed how one of the most important presuppositions of the
Kochen-Specker contradiction can be related to the logical notion of
truth functionality. Weaker versions of admissible valuations also
fail to give a truth-functional system. This is the case, for
example, of the valuations contained in the set $NS3$. Given that
valuations in a semantics-based in non-deterministic matrices are
not, in general, truth-functional, they could provide an interesting
way of describing those formal aspects of quantum theory related to
Kochen--Specker-like contextuality. Thus, given that quantum states
cannot be interpreted in terms of classical (deterministic)
valuations (i.e., valuations in $CAV$), in this section we aim to
describe them as valuations of a non-deterministic semantics. It
turns out that there are several ways to associate non-deterministic
matrices to the quantum formalism. Furthermore, we show that quantum
states can be described as valuations associated with a very
particular form of non-deterministic truth tables.

\subsection{Construction of the N-Matrices for the Quantum Formalism}

In this section, we build the canonical N-matrices of the quantum
formalism. We use the lattice of propositions
$\mathcal{L}(\mathcal{H})$ appearing in \eqref{e:nonkolmogorov} and
the physical constraints imposed by the properties of quantum
states, in such a way that the valuations defined by our matrices
are exactly those given by quantum states. Let $V=[0,1]$ and
$D=\{1\}$. A proposition will be true if and only if its valuation
yields the value 1, and it is false for any other value (this is
connected to the standard quantum logical notion of truth; see
discussion in \cite{piron}). In order to build the matrices, we take
into account first that for every state $\mu$, whenever $P\bot Q$,
we have $\mu(P\vee Q)=\mu(P)+\mu(Q)$ (this follows from
\eqref{e:nonkolmogorov}). Let us now study the disjunction function
$\widetilde{\vee}: V\times V \rightarrow \mathcal{P}(V)\setminus
\{\emptyset\}$. From the Equation \eqref{e:nonkolmogorov}, it is
easy to check that $\max(\mu(P),\mu(Q))\leq \mu(P\vee Q)\leq 1$. In
terms of valuations (denoted generically by $v$), this can be
written as
$$\max(v(P),v(Q))\leq v(P\vee Q)\leq 1.$$

In this way, the natural candidate for the disjunction matrix is
given by
\begin{equation}\label{e:QuantumVeeTable}
    \begin{tabular}{c|c|c|c}
  $ $ & $P$  & $Q$ & $\widetilde{\vee}$  \\ \hline
   if $P\perp Q$  & $a$ & $b$ & $\{a+b\}$\\
if $P\not \perp Q$ & $a$ & $b$ & $[\max(a,b);1]$
 \end{tabular}
\end{equation}

Let us now turn to the functions associated to the conjunction. They
have to be of the form $\widetilde{\wedge}: V\times V\rightarrow
\mathcal{P}(V)\setminus\{\emptyset\}$. Proceeding as before, we
obtain that valuations should satisfy
$$v(P\wedge Q)\leq \min(v(P), v(Q)).$$

Thus, whenever a valuation $v(\cdots)$ assigns $ v(P)=a$ and
$v(Q)=b$  ($a,b\in V$), it is reasonable to define the following
N-matrix:
\begin{equation}\label{e:QuantumWedgeTable}
    \begin{tabular}{c|c|c|c}
  $ $ & $P$  & $Q$ & $\widetilde{\wedge}$  \\ \hline
if $P\not \perp Q$ & $a$ & $b$ & $[0,\min(a,b)]$\\
\hbox{if} $P\perp Q$ & $a$ & $b$ & $\{0\}. $ \\
 \end{tabular}
\end{equation}

Notice that in the tables (\ref{e:QuantumVeeTable}) and
(\ref{e:QuantumWedgeTable}) 
the {connectives} ``$\vee$'' and ``$\wedge$'' are defined by parts,
in the sense that we distinguish between their restrictions to the
case of orthogonal vs. non-orthogonal input propositions. Formally
speaking, this means that each Table describes the {functions}
associated with two partially defined connectives (each line of each
Table defines a function associated with a partially
defined~connective).

It remains to give the table for the negation $ \widetilde{\neg} : V
\rightarrow \mathcal{P}(V) \setminus \{\emptyset\}$. The most
natural candidate compatible with the properties of quantum states
\eqref{e:nonkolmogorov} is given by $\widetilde{\neg}(a)=\{1-a\}$,
and then
\begin{equation}\label{e:NegTableQuan}
    \begin{tabular}{c|c}
 $P$ & $\widetilde{\neg}$  \\ \hline
    $a$ & $\{1-a\}$\\
 \end{tabular}
\end{equation}

This is a deterministic negation, in the sense that its
interpretation set is a singleton, which is a function of $a$.

The above three tables (considered together) impose the restrictions
for all possible valuations. A closer look at them reveals that {for
finite-dimensional models} they contain the very conditions of
Gleason's theorem. Thus, the only valuations that satisfy them (for
finite-dimensional models) are those defined by quantum states. Let
us see how this is so. First, notice that any quantum state will
satisfy the three tables. Thus, the set of quantum states is
contained in the set of valuations defined by the above tables.
Conversely, suppose that a given valuation $v$ satisfies the three
tables. As a valuation, it will take values in the interval $[0,1]$.
In order to see that condition $1$ of \eqref{e:nonkolmogorov} is
valid, let us first take two arbitrary orthogonal one-dimensional
projections $P$ and $Q$. Since they are orthogonal, using the
truth-table of the conjunction, we have $v(P\wedge Q)=0$. On the
other hand, given that $P\wedge Q=\textbf{0}$ (since they are
orthogonal), we conclude that $v(\textbf{0})=v(P\wedge Q)=0$, which
is condition $1$ of \eqref{e:nonkolmogorov}. It is also important to
remark that if $P\leq Q$, then $P\vee Q=Q$ and, using
\ref{e:QuantumVeeTable} for the case $P\not \perp Q$ (when $P\neq
0$), it follows that $v(Q)=v(P\vee Q)\geq \max(v(P),v(Q))\geq v(P)$.
Thus, it easily follows that any valuation compatible with the above
tables will be \emph{order preserving}. When $P\perp Q$, we have
$P\leq Q^{\perp}$, and then, $v(P)\leq v(Q^{\perp})$. Using
\ref{e:NegTableQuan}, this can be rewritten as $v(P)\leq 1-v(Q)$.
Thus, it follows that for $P\perp Q$, $v(P)+v(Q)\leq 1$ (notice
that, in the first line of \ref{e:QuantumVeeTable}, we could have
set $\{\min(a+b,1)\}$ instead of the simpler expression $\{a+b\}$;
but since $v(P)+v(Q)\leq 1$ for all $v$, this is not necessary).
Whenever two orthogonal projections are given, due to the
truth-table of the disjunction, we have $v(P\vee Q)=v(P)+v(Q)$. For
finite-dimensional models, this is equivalent to condition $3$ in
\eqref{e:nonkolmogorov}. Condition $2$ of \eqref{e:nonkolmogorov} is
automatically satisfied due to the truth table of the negation.
Thus, we have proved that any valuation satisfying the three tables
also satisfies the axioms of quantum states (for finite-dimensional
models), and then, it has to be a quantum state. In other words, we
have that, for finite-dimensional models, the only valuations that
satisfy the above non-deterministic truth-tables are the quantum
states (i.e., those states defined by density matrices).

For infinite-dimensional models of quantum systems, one could still
try to impose a $\sigma$-additivity condition (such as the one
appearing in (\ref{e:nonkolmogorov})), and apply again Gleason's
theorem to obtain quantum states. This involves the use of a
condition on a denumerable set of propositions. We will study this
possibility in a future work. It is also important to remark that
there are generalized versions of Gleason's {theorem}
\cite{Hamhalter} that can be used to study additive (and not
necessarily $\sigma$-additive) measures, and could be an interesting
subject of study in future works.

One last thing remains. According to the above-defined tables, we
may ask: are the above tables truth-functional with regard to the
quantum logical connectives? To begin with, notice that the negation
table is classical. More concretely: is it true that for every two
propositions $P$ and $Q$, and every quantum logical connective
$\diamond$, if the probabilities assigned to $P$ and are $a$ and
$b$, respectively, the value of the probability of $P\diamond Q$ is
determined by $a$ and $b$? This question is tricky, because, even if
the valuations are valued into sets with more than one element,
Gleason's theorem could impose, in principle, restrictions in such a
way that all states that assign probabilities $a$ and $b$ to $P$ and
$Q$, respectively, assign the same value for the composed
proposition $P\diamond Q$. This is explicitly the case when $P\perp
Q$: for all $\rho\in\mathcal{S}(\mathcal{H})$, if $\mbox{tr}(\rho
P)=a$ and $\mbox{tr}(\rho P)=b$, we have that $\mbox{tr}(\rho(P\vee
Q))=a+b$ and $\mbox{tr}(\rho(P\wedge Q))=0$. It turns out that this
is not the case when $P\not\perp Q$, as the following examples~show.

Consider first a four dimensional quantum model and the basis
$\{|a\rangle,|b\rangle,|c\rangle,|d\rangle\}$. Consider the
propositions defined by the projection operators $P=|a\rangle\langle
a|+|b\rangle\langle b|$ and $Q=|b\rangle\langle b|+|c\rangle \langle
c|$. Clearly, $R:=P\wedge Q=|b\rangle\langle b|$ and $P\not\perp Q$.
We chose for simplicity $\alpha,\beta,\gamma,\delta\in\mathbb{R}$
and consider the state
$|\psi\rangle=\sqrt{\alpha}|a\rangle+\sqrt{\beta}|b\rangle+\sqrt{\gamma}|c\rangle+\sqrt{\delta}|d\rangle$.
Thus, the probabilities of the elements of the basis are given by
$\alpha$, $\beta$, $\delta$ and $\gamma$, respectively, and the
normalization condition reads $\alpha+\beta+\gamma+\delta=1$. A
simple calculation yields that the probability of $P$ is
$p_{\psi}(P)=\mbox{tr}(|\psi\rangle\langle\psi|P)=\alpha+\beta$, the
probability of $Q$ is
$p_{\psi}(Q)=\mbox{tr}(|\psi\rangle\langle\psi|Q)=\beta+\gamma$ and
the probability of $R$ is $p_{\psi}(P\wedge
Q)=\mbox{tr}(|\psi\rangle\langle\psi|(P\wedge Q))=\beta$. Now, chose
$0<\varepsilon< \alpha,\beta,\gamma,\delta$, and consider a new
quantum state defined by
$|\psi_{\varepsilon}\rangle=(\sqrt{\alpha+\varepsilon}|a\rangle+\sqrt{\beta-\varepsilon}|b\rangle+\sqrt{\gamma+\varepsilon}|c\rangle+\sqrt{\delta-\varepsilon}|d\rangle$
(the reader can easily check that the normalization is correct).
Now, we have that the probability of $P$ is
$p_{\psi_{\varepsilon}}(P)=\mbox{tr}(|\psi_{\varepsilon}\rangle\langle\psi_{\varepsilon}|P)=\alpha+\varepsilon+\beta-\varepsilon=\alpha+\beta$,
the probability of $Q$ is
$p_{\psi_{\varepsilon}}(Q)=\mbox{tr}(|\psi_{\varepsilon}\rangle\langle\psi_{\varepsilon}|Q)=\beta-\varepsilon+\gamma+\varepsilon=\beta+\gamma$
and the probability of $R$ is $p_{\psi_{\varepsilon}}(P\wedge
Q))=\mbox{tr}(|\psi_{\varepsilon}\rangle\langle\psi_{\varepsilon}|(P\wedge
Q)=\beta-\varepsilon\neq \beta$. Thus, we have two {different}
states that assign the same probabilities to $P$ and $Q$, but
{different} values to $P\wedge Q$.

With the same notation as in the previous example, we have that
$S:=P\vee Q=|a\rangle\langle a|+|b\rangle\langle b|+|c\rangle\langle
c|$. Again, we have $P\not\perp Q$, $p_{\psi}(P)=\alpha+\beta$ and
$p_{\psi}(Q)=\beta+\gamma$. For the disjunction, we now have
$p_{\psi}(P\vee Q)=\mbox{tr}(|\psi\rangle\langle\psi|(P\vee
Q))=\alpha+\beta+\gamma$. Computing the probabilities for state
$|\psi_{\varepsilon}\rangle$, we obtain again
$p_{\psi_{\varepsilon}}(P)=\alpha+\beta$ and
$p_{\psi_{\varepsilon}}(Q)=\beta+\gamma$. The probability of the
disjunction is given by $p_{\psi_{\varepsilon}}(P\vee
Q)=\mbox{tr}(|\psi_{\varepsilon}\rangle\langle\psi_{\varepsilon}|(P\vee
Q)=\alpha+\varepsilon+\beta-\varepsilon+\gamma+\varepsilon=\alpha+\beta+\gamma+\varepsilon\neq
\alpha+\beta+\gamma$. Thus, we have two {different} states that
assign the same probabilities to $P$ and $Q$, but {different} values
to $P\vee Q$.

These examples show that the truth tables defined above define a
strictly non-deterministic semantics: the valuations which are
compatible with those tables are {dynamic}. Are they static? The
following example shows that this is not the case.

Consider a three dimensional Hilbert space with a basis
$\{|a\rangle,|b\rangle,|c\rangle\}$. Define $P=|a\rangle\langle a|$
and $Q=|\varphi\rangle\langle\varphi|$ (where
$|\varphi\rangle=\frac{1}{\sqrt{2}}(|a\rangle+|b\rangle)$)). The
conjunction is given by $P\vee Q=|a\rangle\langle
a|+|b\rangle\langle b|$ (due to the fact that they are two linearly
independent vectors, they define a closed subspace of dimension
$2$). Consider the state $|\phi\rangle=|b\rangle$. Thus, we have
$p_{\phi}(P)=\mbox{tr}(|b\rangle\langle b|a\rangle\langle a|)=0$,
$p_{\phi}(Q)=\frac{1}{2}$ and $p_{\phi}(P\vee
Q)=\mbox{tr}(|b\rangle\langle b|(P\vee
Q))=\mbox{tr}(|b\rangle\langle b|(|a\rangle\langle
a|+|b\rangle\langle b|))=1$. Consider now $P'=|c\rangle\langle c|$
and $Q'=Q$. We again have $p_{\phi}(P')=\mbox{tr}(|b\rangle\langle
b|c\rangle\langle c|)=0$ and $p_{\phi}(Q')=\frac{1}{2}$. The
disjunction is given by $P'\vee Q'=|\varphi\rangle\langle
\varphi|+|c\rangle\langle c|$ (we are using that $|c\rangle$ and
$|\varphi\rangle$ are orthogonal). Now, $p_{\phi}(P'\vee
Q')=\mbox{tr}(|b\rangle\langle b|(|\varphi\rangle\langle
\varphi|+|c\rangle\langle c|))=\mbox{tr}(|b\rangle\langle
b|\varphi\rangle\langle \varphi|)+\mbox{tr}(|b\rangle\langle
b|c\rangle\langle c|=\frac{1}{2}+0=\frac{1}{2}\neq 1$. Thus, we have
obtained that, given two pairs of propositions, $P$ and $Q$, and
$P'$ and $Q'$, there is a valuation
$v_{\phi}(...)=\mbox{tr}(|\phi\rangle\langle\phi|(...))$ (induced by
the quantum state $|\phi\rangle$), that satisfies
$v_{\phi}(P)=v_{\phi}(P')$, $v_{\phi}(Q)=v_{\phi}(Q')$, but
$v_{\phi}(P\vee Q)\neq v_{\phi}(P'\vee Q')$. Thus, in general, the
valuations will not be static. Thus, the valuations defined by
quantum states {will not be static} in general.

In Table \ref{tab:table2}, we summarize the differences between the
classical and quantum cases.

\begin{table}[h!]
  \centering
    \caption{Table comparing the different valuations that can be defined on classical vs. quantum propositional systems.}
    \label{tab:table2}
    \begin{tabular}{|l|c|r|} 
      \hline
       & \textbf{Classical systems} & \textbf{Quantum systems} \\
       \hline
      Lattice & Boolean Algebra & Projections lattice \\
        \hline
      Truth-tables & Admit deterministic matrices & Only proper N-matrices\\
        \hline
      Truth-Values & Admit valuations in $\{0,1\}$ & Only valuations in $[0,1]$\\
        \hline
      Truth-Functional & Yes (for deterministic states) & No\\
        \hline
      Satisfy Adequacy & Yes (for deterministic states)  & No\\
        \hline
    \end{tabular}
\end{table}

\subsection{The General Case}

In this section, we turn into the non-deterministic matrices for
generalized probabilistic models, whose propositional structures are
defined by arbitrary complete orthomodular lattices and states are
defined by Equation \eqref{e:generalizedstates}. We proceed in a
similar way to that of the quantum case, but it is important to take
into account that now (a) Gleason's theorem will no longer be
available in many models, and (b) the difference between additivity
and $\sigma$-additivity imposes a great restriction if one wants to
link valuations with states (see \cite{Hamhalter} for
generalizations of Gleason's theorem and the discussion about the
difference between additivity and $\sigma$-additivity). In the
general case, it will be possible to affirm that every state defines
a valuation, but there will exist valuations which are no states.
Furthermore, if the lattice admits no states
\cite{Greechie-NoStates}, then the matrices that we define will
admit no valuations at all.

It is also very important to remark that the N-matrices introduced
below, work well when the lattices are Boolean algebras. This means
that classical (Kolmogorovian) probabilistic models (defined by
Equation \eqref{e:kolmogorovian}) {also} fall into our scheme.
Indeed, following a similar procedure to the one of the preceding
section, it is possible to check that the N-matrices associated with
a classical probabilistic model are strictly non-deterministic. A
similar conclusion can be reached from the point of view of
multi-valued logics \cite{Pykacz-Book}. Alike the quantum case,
classical probabilistic models {always} admit global classical
valuations whose range is {equal} to the set $\{0,1\}$ (i.e., they
admit valuations obeying Equations
\eqref{e:QTruthF}--\eqref{e:QNeg}). If the requirements of physics
are satisfied, due to the KS and Gleason's theorems, the range of
global valuations associated with the N-matrices associated with
quantum systems {cannot be equal} to $\{0,1\}$. A similar
observation holds for more general probabilistic models, provided
that they are contextual enough and they admit states.

Let us first build the interpretation set for the conjunction
function $\widetilde{\wedge}: V\times V\rightarrow
\mathcal{P}(V)\setminus\{\emptyset\}$. Using Equation
\eqref{e:generalizedstates}, it easily follows that $\mu(p\wedge
q)\leq \min(\mu(p),\mu(q))$. Remember also that, in the general
setting, $p\perp q$ iff $p\leq q^{\perp}$ (or equivalently, $q\leq
p^{\perp}$), where $(...)^{\perp}$ is the orthocomplementation in
$\mathcal{L}$ (see \cite{kalm83}, Chapter $1$). Thus, whenever a
valuation $v(\cdots)$ assigns $ v(p)=a$ and $v(q)=b$ ($a,b\in V$),
the natural truth table for the conjunction can be given~by:
\begin{equation}\label{e:GenWedgeTable}
    \begin{tabular}{c|c|c|c}
  $ $ & $p$  & $q$ & $\widetilde{\wedge}$  \\ \hline
 if $p\not\perp q$ & $a$ & $b$ & $[0,\min(a,b)]$\\
 \hbox{if} $p\perp q$ & $a$ & $b$ & $\{0\} $
 \end{tabular}
\end{equation}

Let us now study the disjunction function $ \widetilde{\vee}:
V\times V \rightarrow \mathcal{P}(V)\setminus \{\emptyset\}$. Using
\eqref{e:generalizedstates}, we obtain $\max(\mu(p),\mu(q))\leq
\mu(p\vee q)\leq 1$ and $\max(v(p),v(q))\leq v(p\vee q)\leq 1$.
Thus, the truth-table is given~by:
\begin{equation}\label{e:GenVeeTable}
    \begin{tabular}{c|c|c|c}
  $ $ & $p$  & $q$ & $\widetilde{\vee}$  \\ \hline
   If $p\bot q$  & $a$ & $b$ & $\{a+b\}$\\
if $p\not \perp q$ & $a$ & $b$ & $[\max(a,b);1]$
 \end{tabular}
\end{equation}

In the general setting, the negation of a proposition $a$ is
represented by its orthogonal complement ``$a^{\bot}$'' in the
lattice. In order to impose restrictions on valuations, we use
\eqref{e:generalizedstates} (in the following section, we will
consider more general possibilities). Thus, we define
$\widetilde{\neg}(a)=\{1-a\}$ and obtain:
\begin{equation}\label{e:NegTableGen}
    \begin{tabular}{c|c}
 $p$ & $\widetilde{\neg}$  \\ \hline
    $a$ & $\{1-a\}$\\
 \end{tabular}
\end{equation}

\section{Other Logical Aspects of Our Construction}\label{s:LogicalConsequence}

In this section, we turn into other logical aspects of the quantum
N-matrices. We start by discussing the notion of adequacy and
introduce variants of the deterministic negation defined by
(\ref{e:NegTableQuan}). Next, we discuss how they behave under the
application of double negation. Finally, we discuss the notion of
logical consequence defined by the quantum N-matrices. As is well
known, the inclusion relationship between projection operators is
not---from a logical point of view---a true implication. By
constructing a suitable N-matrix semantics for the quantum
formalism, we would obtain the benefits of the notion of logical
consequence developed by Avron and Zamansky \cite{Avron-Zamansky}.
One of the most important advantages of the N-matrix system is that,
given that it is usually represented by finite matrices, it is
possible to prove its decidability. In our approach---in order to
obtain a closer connection with the quantum formalism---we assumed
that $V$ is non-denumerable. We show below that it is always
possible to reduce the cardinality of $V$ to that of a denumerable
set, without affecting the set of theorems. This is directly related
to the definitions of F-expansion and
refinement~\eqref{d:Refinement}.

\subsection{Quantum N-Matrices and Adequacy}

Let us now turn to the notion of adequacy of N-matrices:

\begin{definition}\label{d:AvronCriterion}
Let $M=\langle V,D,O\rangle$ be an N-matrix for a language which
includes the positive fragment of the classical logic ($LK^{+}$). We
say that $M$ is {adequate} for this language, in case that the
following conditions are~satisfied:

\begin{enumerate}
\item $\mathbf{\widetilde{\wedge}}$:  $$\mbox{If}\,\,\, a\in D\,\, \mbox{and}\,\, b\in D,\,\,\mbox{then}\,\,
a\widetilde{\wedge}b \subseteq D$$
$$\hbox{If}\,\,\, a\not\in D,\,\, \hbox{then}\,\, a\widetilde{\wedge}b \subseteq V\setminus{D} $$
$$\hbox{If}\,\,\, b\not\in D, \,\,\hbox{then}\,\, a\widetilde{\wedge}b \subseteq V\setminus{D} $$

\item $\mathbf{\widetilde{\vee}}$: $$\hbox{If}\,\,\, a\in D  ,\,\,\hbox{then}\,\,
a\widetilde{\vee}b \subseteq D $$
$$\hbox{If}\,\,\, b\in D  ,\,\,\hbox{then}\,\, a\widetilde{\vee}b \subseteq D $$
$$\hbox{If}\,\,\, a\not \in D  \,\,\hbox{and}\,\, b\not \in D,\,\,\hbox{then}\,\, a\widetilde{\vee}b \subseteq V\setminus{D} $$ \\

\item $\mathbf{\widetilde{\supset}}$: $$ \hbox{If}\,\,\, a\not\in D, \,\,\hbox{then}\,\, a
\widetilde{\supset}b\subseteq D$$
$$ \hbox{If}\,\,\, b\in D, \,\,\hbox{then}\,\, a \widetilde{\supset}b\subseteq D$$
$$\hbox{If}\,\,\, a\in D \,\,\hbox{and}\,\, b\not \in D,\,\,\hbox{then}\,\, a\widetilde{\supset}b \subseteq V\setminus{D}$$
\end{enumerate}
\end{definition}

In this section we briefly discuss definition \ref{d:AvronCriterion}
in the context of quantum generalized N-matrices with finite
precision measurements. By this, we mean a situation in which it is
not possible to determine if a proposition is true, but it is
possible to assure that it is included within a certain interval
around $1$. In order to do this,
we can now take the same tables defined by \eqref{e:QuantumVeeTable}--\eqref{e:NegTableQuan} 
(or equivalently, \eqref{e:GenVeeTable}, \eqref{e:GenWedgeTable} and
\eqref{e:NegTableGen} for the general case), and change the
designated values set to $D=[\alpha,1]$, with $\alpha \in (0,1].$ By
setting $\alpha=1$, we obtain the matrices of the previous sections.

 It is important to remark that with an interpretation set such
as $D=[\alpha,1]$ and tables defined by
\eqref{e:QuantumVeeTable}--\eqref{e:NegTableQuan} (or
\eqref{e:GenVeeTable}, \eqref{e:GenWedgeTable} and
\eqref{e:NegTableGen}), it could be the case that a valuation
selects a non-designated value, even when both input elements are
designated (notice that this is also true for the case $\alpha=1$).
Let us illustrate this with an example. Suppose that $\alpha=1$ and
we prepare a quantum system in a state
$|\psi\rangle=\frac{1}{\sqrt{2}}(|\psi_{1}\rangle+|\psi_{2}\rangle)$
with $\langle\psi_{1}|\psi_{2}\rangle=0$. Thus, the valuation
associated to the state $\rho_{\psi}=|\psi\rangle\langle\psi|$,
assigns non-designated values to the propositions
$|\psi_{1}\rangle\langle\psi_{1}|$ and
$|\psi_{2}\rangle\langle\psi_{2}|$, while it assigns a designated
value to the proposition
$|\psi_{1}\rangle\langle\psi_{1}|+|\psi_{2}\rangle\langle\psi_{2}|$
(associated to the disjunction $|\psi_{1}\rangle\langle\psi_{1}|$
and $|\psi_{2}\rangle\langle\psi_{2}|$). This makes our matrix
non-adequate because it violates condition $2$ of definition
\ref{d:AvronCriterion}. Thus, in general, quantum states will not
satisfy adequacy. Adequacy is not a requirement for physics, but it
could be of interest to logicians. For example, it is possible to
use the criterion of definition \ref{d:AvronCriterion} to give
unicity proofs for matrices given a certain language and conditions.
Thus, let us see what can we do in order to obtain an adequate
matrix. We must first restrict the interpretation set of the
conjunction. Notice that, by doing this, we depart from physics,
given that the new valuations may no longer be quantum~states.

 Let us start by defining the interpretation set for the conjunction
$\widetilde{\wedge}:V\times V\rightarrow \mathcal{P}(V)\setminus
\{\emptyset\}$. As before, $a$ and $b$ represent the respective
values for the valuations of $p$ and $q$: $v(p)=a$ and $v(q)=b$. Let
us discuss case by case.

\begin{itemize}
    \item If $a,b \in D$:
    \end{itemize}

Given that $0\leq v(p\wedge q)\leq min(a,b)$, this suggests us to
take the interpretation set for the conjunction for the case where
both values are designated as:
\begin{equation}
a\widetilde{\wedge}b \subseteq [0,min(a,b)]\cap D
\end{equation}
\begin{itemize}
    \item If $a\in D,$ and $b\not \in D $
\end{itemize}

 Proceeding similarly as before, we obtain $v(p\wedge q)\leq min(a,b)$, but now we know which
is the smallest between $a$ and $b$, given that $b$ is not
designated. Thus,
$$0\leq v(p\wedge q)\leq b $$ This suggests the following
interpretation:
\begin{equation}
a\widetilde{\wedge}b \subseteq [0,b] \subseteq V\setminus D.
\end{equation}

 It is easy to check that this case satisfies the adequacy
criterion without the necessity of restricting the set.

The case $a\not \in D$ and $b\in D$ is totally analogous.
\begin{itemize}
    \item If $a,b\not \in D$:
\end{itemize}

Given that $$ 0\leq v(p\wedge q)\leq \min(a,b),\,\, \hbox{thus}$$
\begin{equation}
    a\widetilde{\wedge}b \subseteq [0,\min(a,b)] \subseteq V\setminus D.
\end{equation}

Proceeding in an analogous way, we now obtain the interpretation set
of the disjunction:
$$ \widetilde{\vee}: V^{2}\rightarrow 2^{V}\setminus \{\emptyset\}$$
\begin{itemize}
    \item If $a,b\in D$
\end{itemize}

Given that $p\leq p\vee q $ and $q\leq p\vee q $, then $\mu(p)\leq
\mu(p\vee q)$ and $\mu(q)\leq \mu(p\vee q)$.
$$ a\leq v(p\vee q) \hbox{ and } b\leq v(p\vee q), \hbox{ thus}$$
$$ \max(a,b)\leq v(p\vee q)\leq 1$$
Thus,
\begin{equation}
    a\widetilde{\vee}b \subseteq [\max(a,b),1]\subseteq D
\end{equation}
\begin{itemize}
    \item If $a\in D, b\not \in D \hbox{ or } a\not \in D, b\in D$
\end{itemize}
\begin{equation}
    a\widetilde{\vee}b \subseteq [\max(a,b),1]\subseteq D
\end{equation}
\begin{itemize}
    \item If $a,b\not \in D$
\end{itemize}

 If none of the two terms has a designated value, one possibility is to proceed as in the first case with the conjunction
(by restricting our set in such a way that it satisfies the adequacy
criterium).
$$a\widetilde{\vee}b \subseteq [\max(a,b),1]$$

A valuation for two projections with non-designated values could
give us a designated value. If we want to avoid this, and using
Avron's criterion, the interpretation set should be:
\begin{equation}
   a \widetilde{\vee}b \subseteq [\max(a,b),1]\cap (V\setminus D).
\end{equation}

We will not follow this choice, given that Gleason's theorem imposes
stronger restrictions on valuations.

Let us turn now to the negation. We consider different choices,
alternative to (\ref{e:NegTableGen}). Given that we are working with
orthomodular lattices, we have that any state satisfies
$\mu(p^{\perp})=1- \mu(p)$. In terms of valuations, this condition
reads: $v(p^\perp) =1-v(p)$.
In order to obtain a non-deterministic negation, our first choice is:

Case 1:
\begin{equation}\label{e:Neg1}
\begin{tabular}{c|c }
     $p$ & $\widetilde{\neg}_{1}$    \\  \hline
     $a\in D$ & $[0,1-a]$\\
     $a\not \in D$ & $[1-a,1]$
\end{tabular}
\end{equation}

This case generalizes the standard quantum one (given by
(\ref{e:NegTableQuan})), leaving $1-a$ (deterministic negation)
as a respective bound. 

Case 2: We introduce parameters now. Let us assume that $\alpha \in
(\frac{1}{2},1)$ and define:
\begin{equation}\label{e:Neg2}
\begin{tabular}{c|c }
     $p$ & $\widetilde{\neg}_{2}$    \\  \hline
     $a\in D$ & $[\alpha-\frac{a}{2}(\frac{1-\alpha}{\alpha}),\alpha)$\\ \\
     $a\not \in D$ & $[\alpha,\alpha+\frac{a}{2}(\frac{1-\alpha}{\alpha})]$
\end{tabular}
\end{equation}

Both sets depend now on the value of $\alpha$.

 Finally, by appealing to \eqref{e:NegTableQuan} and \eqref{e:NegTableGen}, we can always define deterministic negations. This is perhaps the more natural choice for the standard quantum case and more general probabilistic models. In this case, independently of whether the value of $a$ is
designated or not, the negation yields $1-a$. If we take $D=\{1\}$
as in the standard quantum logical case, then the negation of a
given designed value would give a non-designed result. The
converse is not necessarily true. 

\subsection{Double Negation}

Now we turn to some problems that could emerge with the behavior of
the double negation. A~more detailed study of the negation and
double negation is left for future work. We consider both, logical
and physical motivations in order to proceed.

Given its relation to the orthogonal complement in Hilbert spaces,
it is desirable that the negation satisfies the principle of double
negation. It is obvious that the negations defined in
\eqref{e:NegTableQuan} and \eqref{e:NegTableGen} satisfy this
principle. We now show that, despite of not respecting this
principle strictly, the negation $\widetilde{\neg}_{1}$ (see
(\ref{e:Neg1})) has a very particular behavior that could be related
to the classical limit between logics. Although the negations
defined by \eqref{e:Neg1} and \eqref{e:Neg2} do not behave properly
in this sense, their existence is interesting on its own, given
that, in a different domain, we may need different types of
connectives associated with the negation. As an example, in quantum
circuits, it is possible to define an operation which is the square
root of the negation. It is important to remark that the failure of
double negation is also found in other algebraic structures
associated with the quantum formalism (see for example
\cite{LandsmanQL-2019}).

Let us now analyze the double negation for $\widetilde{\neg}_{1}$.
Applying it twice, we obtain:
$$ v((\neg_{1}(\neg_{1}(p)))\in \widetilde{\neg}_{1}(v(\neg_{1} p))$$
$$v(\neg_{1}p)\in \widetilde{\neg}_{1}(v(p))$$

Let us analyze the different cases in relation to the set of
designated values. We assume that $ 0.5 < \alpha$.
\begin{itemize}
\item If $v(p)=a \in D=[\alpha ,1]$:
\end{itemize}

$$\widetilde{\neg}_{1}(a)= [0,1-a] \hbox{ and } v(\neg_{1} p)\in [0,1-a]$$

Let $b=v(\neg_{1}p)\in [0,1-a]\subseteq (V\setminus D)$. Then:
$$v(\neg_{1}(\neg_{1} p)) \in \widetilde{\neg}_{1}(b)=[1-b,1],$$
given that $b$ is not a designated value. Thus, the double negation
maps designated values to designated values. If we order all
possible values, then:
$$ 0 \leq b \leq 1-a \leq \alpha \leq a \leq 1-b \leq 1 $$

This proves that, after taking two times the negation of a
designated value, the value for the double negation must be chosen
out of a set which is included in the set from which the original
designated value was taken. This means that, when taking
consecutively the double negation for the designated case, the lower
bound of the final interpretation set for the connective is closer
to $\{1\}$. Now we show that this will not happen if we start from a
non-designated value.

\begin{itemize}
    \item  If $v_{(p)}=a\not \in D$
\end{itemize}

$$\widetilde{\neg}_{1}(a)=[1-a,1]$$

Let $b$ be the value that the valuation takes inside this set:
$$b=v_{(\neg_{1} p)}\in [1-a,1]$$

Then, we obtain
$$v(\neg_{1}(\neg_{1} p))\in \widetilde{\neg}_{1}(b)$$

The difference with regard to the previous case is that now $b$
could be a designated value or not, given that inside the interval
$[1-a,1]$, in principle, there could be values of both types. As an
example, if $\alpha= 0.9$ and $a=0.8$, the involved interval would
be $[0.2,1]$, which contains both designed and non-designed values.
Thus, for the double negation, in this case, we have to separate the
interpretation set depending on the type of value taken by $b$.

This behavior was already included in the orthodox quantum logical
treatment, where the negation of a true proposition was false, but
the negation of something false is not necessarily true. Thus, we
obtain:
$$v(\neg_{1} (\neg_{1} p))\in \widetilde{\neg}_{1}(b)$$
with:
\[ \widetilde{\neg}_{1}(b) =  \left\{
\begin{array}{ll}
      [0,1-b] & b\in D\\
      \left[1-b,1\right]  & b\not\in D \\
\end{array}
\right. \]

This means that the application of the double negation many times
not necessarily has as a consequence a concentration of the values
of the valuation around $0$.

A similar analysis can be made for the second negation presented in
this work, and the conclusion---though not identical---goes in the
same direction. The principle of double negation is not satisfied
either.

\subsection{Quantum N-Matrix as a Refinement of an F-Expansion of a Finite N-Matrix}

In this section, we make use of the following definition (presented
in \cite{Avron-Zamansky.Tutorial}):

\begin{definition}\label{d:Refinement}
Let $ M_{1}=\langle V_{1},D_{1},O_{1}\rangle  $ and $ M_{2}=\langle
V_{2},D_{2},O_{2}\rangle$ be N-matrices for $\mathcal{L}$.
\begin{enumerate}

\item $ M_{1}$ is a refinement of $M_{2}$ if $V_{1}\subseteq V_{2}$, $D_{1}=D_{2}\cap V_{ 1}$, and $\widetilde{\diamondsuit}_{M_{1}}(\overline{x})\subseteq \widetilde{\diamondsuit}_{M_{2}}(\overline{x})$ for every n-ary conective $\diamondsuit$ of $L$ and every tuple $\overline{x}\in V^{n}_{1}$.
\item  Let $F$ be a function that
assigns to each $x\in V$ a non-empty set $F(x)$, such that
$F(x_{1})\cap F(x_{2})= \emptyset $  if $x_{1}\neq x_{2}$. The
$F$-expansion of $M_{1}$ is the following N-matrix
$M^{F}_{1}=\langle V_{F},D_{F},O_{F}\rangle $, with $V_{F}= \bigcup
_{x\in V} F(x)$, $D_{F}=\bigcup _{x\in D} F(x)$, and
$\widetilde{\diamondsuit}_{M^{F}_{1}}(y_{1},...,y_{n})=\bigcup_{z\in
\widetilde{\diamondsuit}_{M_{1}}(x_{1},...,x_{n})}F(z) $ whenever
$\diamondsuit$ is an n-ary connective of $L$, and $x_{i}\in V$,
$y_{i}\in F(x_{i})$ for every $1\leq i \leq n$. We say that $M_{2}$
is an expansion of $M_{1}$ if $M_{2}$ is the $F$-expansion of
$M_{1}$ for some function $F$.
\end{enumerate}
\end{definition}

We now show that the N-matrix constructed for the orthomodular
lattice of projection operators for the case of $D=1$ is a
particular refinement of an F-expansion of a finite N-matrix. We
will not consider the case with more designated values. In order to
reach this aim, we must give first some definitions with regard to
expansions and refinements. For a more detailed treatment of the
techniques used in this section (and proofs and propositions), we
refer the reader to \cite{Avron fuzzy}.

We say that $M_{2}$ is a {simple refinement} of $M_{1}$, if it is a
refinement (definition \eqref{d:Refinement}) and it satisfies
$V_{1}= V_{2}$. Given a function $F$, let $Im(F)$ and $Dom(F)$
denote the image and domain of $F$, respectively. For every
expansion function $F$ and $y\in \bigcup Im(F)$, we denote by
$\widetilde{F}[y]$ the unique element $x\in Dom(F)$ such that $y\in
F(x)$.

\begin{definition}
Let $M_{1}= \langle V_{1}, D_{1}, O_{1}\rangle $ and $M_{2}= \langle V_{2}, D_{2}, O_{2}\rangle $ be N-matrices and F an expansion function for $M_{1}$. We say $M_{2}$ is an F-rexpansion of $M_{1}$ if it is a refinement of the F-expansion of $M_{1}$. It is called:\\
\begin{enumerate}
    \item simple if it is a simple refinement of the F-expansion of $M_{1}$.
    \item preserving if $F(x)\cap V_{2}\not = \emptyset$ for every $x\in V_{1}$.
    \item strongly preserving if it is preserving, and for every $x_{1},..., x_{n}\in V_{2}$, $\diamondsuit \in \diamondsuit^{n}_{\mathcal{L}}$ and $y\in \widetilde{\diamondsuit}_{1}(\widetilde{F}[x_{1}],...,\widetilde{F}[x_{n}])$, it holds that the set $F(y)\cap \widetilde{\diamondsuit}_{2}(x_{1},..., x_{n})$ is not empty.
\end{enumerate}
\end{definition}

Loosely speaking, being a preserving rexpansion amounts to keeping
at least one ``copy'' of every original truth-value. Being strongly
preserving means that this property holds not only for the set of
truth-values, but also for the interpretation of the connectives.

\begin{prop}
Every simple rexpansion is preserving, every expansion is a strongly
preserving rexpansion, and every preserving rexpansion of a matrix
is strongly preserving.
\end{prop}

The proof can be found in \cite{Avron fuzzy}.

\begin{prop}\label{p:PropA}
The N-matrix $M_{2}=\langle V_{2},D_{2},O_{2} \rangle$ is a
rexpansion of the N-matrix $M_{1}=\langle V_{1},D_{1},O_{1} \rangle$
iff there is a function $f:V_{2}\rightarrow  V_{1}$ such that:

\begin{enumerate}
    \item For every $x\in V_{2}$, $x\in D_{2}$ iff $f(x)\in D_{1}$.
    \item For every $x_{1},..., x_{n}\in V_{2}$ and  $y \in \widetilde{\diamondsuit}_{2}(x_{1},..., x_{n}) $, it holds that $f(y)\in \widetilde{\diamondsuit}_{1}(f(x_{1}),...,f(x_{n}))$.
\end{enumerate}
\end{prop}

\begin{prop}\label{p:PropB}
If $M_{2}$ is a rexpansion of $M_{1}$ then $\vdash _{M_{1}}
\subseteq  \vdash _{M_{2}}$. Moroever, if $M_{2}$ is strongly
preserving, then $\vdash _{M_{1}} =  \vdash _{M_{2}}$.
\end{prop}

Now we proceed to find a finite N-matrix of which our quantum
N-matrix is an expansion. This N-matrix will not be unique, given
that there exist infinitely many rexpansions for a given N-matrix,
and each matrix can be the rexpansion of different N-matrices. Each
one of these rexpansions is compromised with different expansion
functions and different degrees of refinement. Once one of these
N-matrixes is found, it is possible to use Proposition
\eqref{p:PropB} in order to relate the sets of theorems.

Let $V_{2}=[0,1]$,  $V_{1}=\{t,T,F\}$, $D_{2}= \{1\}$  and $f:V_{2}\rightarrow V_{1}$, such that \\
$$f(1)=t ;\hspace{3mm} f(0)= F;\hspace{3mm} f(\alpha)= T,\,\, \alpha\in (0,1)$$

 In this case, we aim to find a finite N-matrix of three
values, in such a way that the quantum N-matrix be its expansion. We
then propose an N-matrix of two values.

Applying item 1 of proposition \eqref{p:PropA}: $x \in \{1\}$ iff
$f(x)\in D_{1}$ $ \Rightarrow D_{1}=\{t\}$.

Now we apply item 2 of Proposition \eqref{p:PropA} in order to find
the interpretation set for each connective.
$$\forall x_{1}, x_{2}\in [0,1],\,\, y \in \widetilde{\vee}_{Q}(x_{1},x_{2})= [\max(x_{1},x_{2}), 1]\Rightarrow f(y)\in
\widetilde{\vee}_{1}(f(x_{1}),f(x_{2}))$$

If $x_{1}=x_{2}=0$, then $y\in [0,1] \Rightarrow f(y)\in
\widetilde{\vee}_{1}(f(0),f(0))$. Thus,
\begin{equation}
    \widetilde{\vee}_{1}(F,F)=\{t,T,F\}
\end{equation}

If $x_{1}=x_{2}=1$,
$$y\in \widetilde{\vee}_{Q}(1,1)=[\max(1,1),1]=\{1\}\Rightarrow f(1)\in \widetilde{\vee}_{1}(t,t).$$

Then,
\begin{equation}
    \widetilde{\vee}_{1}(t,t)=\{t\}
\end{equation}

If $x_{1}=0, x_{2}=\alpha ; \alpha\in (0,1)$, $y\in [\alpha,1]
\Rightarrow f(y)\in \widetilde{\vee}_{1}(F,T)$.
\begin{equation}
    \widetilde{\vee}_{1}(F,T)=\{t,T\}
\end{equation}

By following this procedure, it is possible to find all the elements
of the interpretation set of the conjunction. The following table
resumes all the results of a possible candidate:
\begin{equation}
    \begin{tabular}{c|c|c|c }
     $\widetilde{\vee}_{1} $ & $ t$ & $T$ & $F $   \\  \hline
    $t$ & $\{t\}$ &$\{t\} $ & $\{t\} $\\
     $T$ &        $\{t\}$ & $\{t,T\} $ &$ \{t,T\}$  \\
     $F$ & $ \{t\}$ & $\{t,T\}$ &$\{t,T,F\}$
\end{tabular}
\end{equation}

It is important to remark that the interpretation set for the
disjunction is one of several possibilities. That in this case, we
are not imposing the constrains related to Gleason's theorem. In the
same way, it is possible to find the set corresponding to the
conjunction. The results are shown in the following table:
\begin{equation}
    \begin{tabular}{c|c|c|c }
     $\widetilde{\wedge}_{1} $ & $ t$ & $T$ & $F $   \\  \hline
    $t$ & $\{t,T,F\}$ &$\{F,T\} $ & $\{F\} $\\
     $T$ &        $\{T,F\}$ & $\{T,F\} $ &$ \{F\}$  \\
     $F$ & $ \{F\}$ & $\{F\}$ &$\{F\}$
\end{tabular}
\end{equation}

For the negation, we have:
\begin{equation}
    \begin{tabular}{c|c|c|c }
     $  $ & $ t$ & $T$ & $F $   \\  \hline
    $\widetilde{\neg}_{1}$ & $\{F\}$ &$\{T\} $ & $\{t\} $
\end{tabular}
\end{equation}

For the above table we have taken the standard negation associated
to the lattice. \\ If $V_{1}=\{t,F\}$ is chosen as the initial set
(instead of $V_{1}=\{t,T,F\}$), it is possible to proceed as
follows:
$$f:V_{2}\rightarrow V_{1}$$
such that $f(1)=t$, $f(\alpha)= F$ and $\alpha\in [0,1)$. Proceeding
in an analogous way as before, we arrive at the following table:
\begin{equation}
    \begin{tabular}{c|c|c }
     $\widetilde{\vee}_{1} $ & $ t$  & $F $   \\  \hline
    $t$ & $\{t\}$ &$\{t\} $  \\
     $F$ & $ \{t\}$ & $\{t,F\}$
\end{tabular}
\end{equation}
 and a similar procedure can be applied to the rest of the
connectives.

\section{Conclusions}\label{s:Conclusions}

In this work we have seen that:

\begin{itemize}
\item There are several ways in which one can affirm that the quantum formalism does not obey truth functionality.
\item The set of projection operators admits N-matrices, and thus, the N-matrices formalism can be adapted to quantum mechanics.
\item Each quantum state can be interpreted as a
valuation associated to a non-deterministic semantics. Indeed, the
set of quantum states can be \textit{characterized} as being
equivalent to the set of valuations defined by the N-matrices that
we propose in section \ref{s:KSandQuantum}. We have proved that
quantum states, considered as valuations, are, in general, dynamic
and non-static. We have provided a similar analysis for generalized
probabilistic models.
\item There exist different candidates for non-deterministic
semantics which are compatible with the quantum formalism. We have
studied different examples.
\item It is possible to give a notion of a logical consequence associated to non-deterministic
semantics in the quantum formalism (a study that should be extended,
of course, in future work).
\end{itemize}

We think that the constructions presented here can open the door to
interesting questions in both, the fields of quantum mechanics and
logic. On the physics side, it opens the door to studying axioms for
generalized probabilistic systems using logical axioms. On the
logical side, it gives place to a new model of N-matrices with
possible physical applications.


\end{document}